# Lévy Noise-Induced Stochastic Resonance in a Bistable System

Yong Xu[†1], Juanjuan Li[1], Jing Feng[1], Huiqing Zhang[1], Wei Xu[1] and Jinqiao Duan[2]

1.  *Department of Applied Mathematics, Northwestern Polytechnical University, Xi'an 710072, China*

2.  *Institute for Pure and Applied Mathematics, UCLA, Los Angeles, CA 90095, USA & Department of Applied Mathematics, Illinois Institute of Technology, Chicago, IL 60616, USA*

*Abstract:*

Stochastic resonance phenomenon induced by non-Gaussian Lévy noise in a second-order bistable system is investigated. The signal-noise-ratio for different parameters is computed by an efficient numerical scheme. The influences of the noise intensity, stability index of Lévy noise and amplitude of external signal on the occurrence of stochastic resonance phenomenon are characterized. This implies that a high amplitude of signal not only enhances the output power spectrum of system but also promotes stochastic resonance, and a proper adjustment of Lévy noise intensity in a certain range enlarges the peak value of output power spectrum which is significant for stochastic resonance. Moreover, with the optimal damping parameter, lowering the stability index leads to larger fluctuations of Lévy noise, and further reduces the chance of the stochastic resonance.

## I.  Introduction

Noise is often thought to have undesirable influences on dynamical systems, although it could help certain systems generate unexpected, ordered patterns, and thus behave in a coherent manner [1, 2]. This intriguing and counterintuitive phenomenon is the so-called stochastic resonance (SR). SR describes a phenomenon of a system in a noisy environment, and at some optimal level of noise intensity, certain characteristics become sensitive to an external periodic stimulus. The concept of SR was put forward by Benzi and collaborators in an attempt to explain the periodicity of the Earth's ice-age cycle [3]. Nicolis independently suggested that SR might rule the periodicity of the recurrent ice ages [4]. SR phenomenon was also observed in a Schmitt trigger circuit and a bidirectional ring laser. This has inspired some recent research on noise-induced effects in various dynamical systems [5,6]. More recently, the SR effect has found wide applications in geophysical, biological, and chemical systems [7-9].

[†] Corresponding author. E-mail: hsux3@nwpu.edu.cn

The ingredients for SR are a nonlinear system, weak periodic signal, and random noise [10]. The effect of SR can be determined by using various system performance measures, including residence time distributions, power spectrum density, signal-to-noise ratio (SNR), spectral power amplification (SPA), input/output cross-correlation measures, and probability of detection [8]. For a given system, the occurrence of SR phenomenon implies that these measures will exhibit a well-marked maximum at a particular noise level [11].

The SR phenomenon in bistable systems with Gaussian noise has been investigated both theoretically and experimentally due to its applications [12-16]. Most authors assume that noise is Gaussian for simplicity [3,12,13-18]. Yet, Gaussian noise is just an ideal case for fluctuations. In fact, in many situations the external noise can be described by distribution of impulses following heavy-tail stable law statistics of infinite variance. Lévy noise, as an important non-Gaussian noise, has heavy-tails, jumps and infinite divisibility, and is considered to be more general than Gaussian noise in dynamical systems. The Lévy noise-driven non-equilibrium systems show interesting physical properties and have been addressed in various physical scenarios exhibiting a super diffusive behavior [19]. Gaussian noise is a special case of Lévy noise. The distribution of Lévy noise exhibits the asymptotic power-law decay, while Gaussian distribution decays exponentially [20].

Lévy noise has been observed in physical, natural, social and complex systems [21]. Dynamical systems excited by Lévy noise have recently attracted the attention of researchers in finance, science and engineering communities [19,22-31], including the periodic SR phenomenon in the presence of Lévy noise [23]. But the SR phenomenon with Lévy noise is considered mostly for first-order dynamical systems rather than second-order systems [19,21-23]. A few authors considered SR for second-order dynamical systems but with Gaussian noise [3,32-34]. So it is worth studying the SR phenomenon induced by non-Gaussian Lévy noise in second-order systems.

In this paper, we consider a prototypical second-order, weakly damped bistable system driven by non-Gaussian Lévy noise together with a periodic input signal, and characterize the corresponding SR phenomenon by using the output signal-to-noise ratio (SNR). We define the SNR as the ratio of the signal power to the noise, while the signal power is obtained by calculating the maximum of the system's output power spectrum with small noise intensity. In addition, we estimate the noise power by taking an average of the output power spectrum in the adjacent signal bins. Our discussions focus on the simulation of SNR with varying noise intensity, stability index of Lévy noise and other parameters, by a proposed numerical method, and then examine the SR effect for various parameters.

This paper is organized as follows. In section II, we introduce a second-order bistable system and Lévy noise. The related potential function and density function of Lévy distribution are obtained in different cases. In section III, we present the simulation results of SNR along with relevant interpretations. We observe that the optimal noise intensity leads to the maximum of SNR, and this maximum value decreases when we weaken the stability index of Lévy noise and the amplitude of the external signal. We end the paper with a discussion in section IV.

## II. A second-order dynamical system with Lévy noise

Consider a second-order system [3,32-34] driven by a Lévy noise

$$\ddot{x} + \gamma \dot{x} + \frac{dV(x)}{dx} = F_1(t) + \eta(t), \tag{1}$$

where $\gamma$ is a damping parameter, $F_1(t) = A\cos(\omega t)$ is an external signal with amplitude $A$ and frequency $\omega$, and $\eta(t)$ denotes the Lévy noise with stability index $\alpha$ $(0 < \alpha \leq 2)$. In fact, $\eta(t)$ is the time derivative of a Lévy process $\zeta(t)$. When $\alpha = 2$, $\eta(t)$ becomes a Gaussian noise. The potential function $V(x)$ in the equation (1) is defined as

$$V(x) = -ax^2/2 + bx^4/4 \ (a > 0, \ b > 0). \tag{2}$$

The function $V(x)$ has two stable fixed points at $x = \pm\sqrt{a/b}$ (i.e. two potential wells centered at $V_+ = \sqrt{a/b}$ and $V_- = -\sqrt{a/b}$, respectively), the height of potential barrier is $\Delta V(x) = a^2/4b$, and the barrier top is located at $x = 0$. We choose the parameters as $a = b = 1$. Fig.1 (a) shows the potential $V(x)$. It is clear that the two potential wells are symmetric and are separated by a barrier whose height is 0.25. When the system is driven by a periodic signal $F_1(t) = A\cos(\omega t)$, the signal will modulate it at each instant time $t$. Fig.1 (b) exhibits the potential function $V_1(x) = V(x) + xF_1(t)$ versus time. The potential barrier of the right and left well can raise and lower successively, which offers a possibility for a particle rolling periodically from one potential well to the other. The motion of the particle between two potential wells reflects the nonlinear dynamical behaviors of the system. It is worth noting that the amplitude $A$ should be smaller than the critical value of bistable system $A_c = \sqrt{4a^3/27b}$ for a weak signal. Fig.1 (c) depicts the potential function $V_1(x)$ with different amplitudes of the periodic signal. It indicates that when a noise is present in the system, the particle may leap from one potential well to another even if $A \ll A_c$, and furthermore, when $A > A_c$, the system's bistability will vanish and reduce a single-well; for example, $A = 0.6 > A_c$, it is apparent that the system's bistability has disappeared.

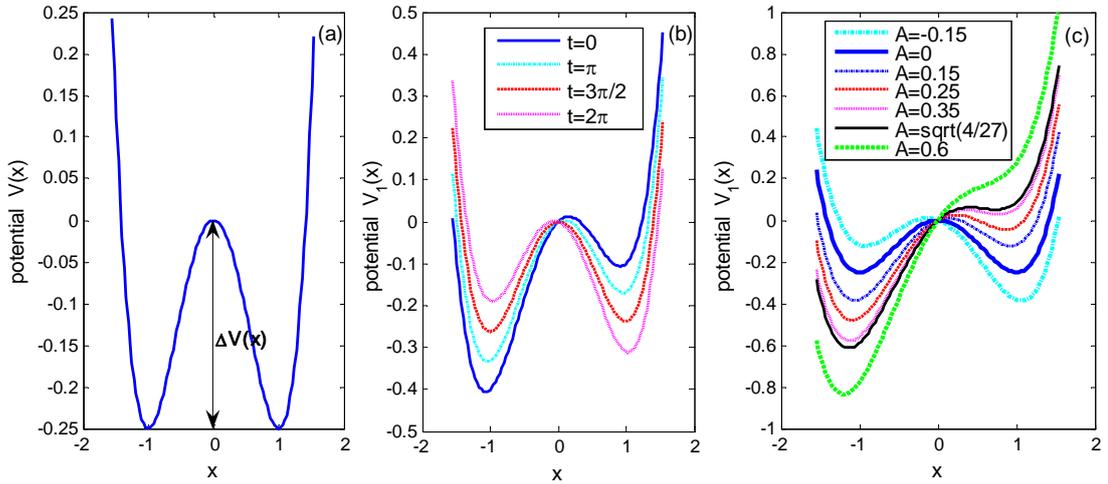

Fig.1. The potential function in different cases: (a) $V(x) = -x^2/2 + x^4/4$. (b) $V_1(x) = V(x) + xA\cos(\omega t), A = 0.15 \neq 0, w = \sqrt{2}/20$, (c) $V_1(x) = V(x) + xA\cos(\omega t)$, $t = \pi/2$.

Assume that $\zeta(t)$ obeys Lévy distribution $L_{\alpha,\beta}(\zeta;\sigma,\mu)$, whose characteristic function is [11]:

$$\Phi(k) = \int_{-\infty}^{+\infty} d\zeta\, e^{ik\zeta} L_{\alpha,\beta}(\zeta;\sigma,\mu).$$

Therefore, for $\alpha \in (0,1) \cup (1,2]$,

$$\Phi(k) = \exp\left[i\mu k - \sigma^\alpha |k|^\alpha \left(1 - i\beta\, sgn(k) \tan\frac{\pi\alpha}{2}\right)\right], \quad (3)$$

and for $\alpha = 1$,

$$\Phi(k) = \exp\left[i\mu k - \sigma |k|\left(1 + i\beta\, sgn(k) \frac{2}{\pi}\ln|k|\right)\right]. \quad (4)$$

Here $\alpha \in (0,2]$ denotes the stability index that describes an asymptotic power law of the Lévy distribution. When $\alpha < 2$, $L_{\alpha,\beta}(\zeta;\sigma,\mu)$ is characterized by a heavy-tail of $|\zeta|^{-(\alpha+1)}$ type with $|\zeta| \gg 1$. The constant $\beta$ ($\beta \in [-1,1]$) is the asymmetry parameter. Moreover, $\sigma$ ($\sigma \in (0,\infty)$) is the scale parameter, $\mu$ ($\mu \in R$) is the mean parameter, and $D = \sigma^\alpha$ represents the noise intensity. In this paper we use the Janicki-Weron algorithm [11,35] to generate the Lévy distribution. For $\alpha \neq 1$, it is simulated as

$$\zeta = D_{\alpha,\beta,\sigma} \frac{\sin[\alpha(U+C_{\alpha,\beta})]}{[\cos(U)]^{1/\alpha}} \left[\frac{\cos[U - \alpha(U+C_{\alpha,\beta})]}{W}\right]^{(1-\alpha)/\alpha} + \mu, \quad (5)$$

with

$$C_{\alpha,\beta} = \frac{\arctan[\beta \tan(\frac{\pi\alpha}{2})]}{\alpha}, \quad (6)$$

and

$$D_{\alpha,\beta,\sigma} = \sigma \left[\cos\left\{\arctan\left[\beta \tan(\frac{\pi\alpha}{2})\right]\right\}\right]^{-1/\alpha}. \quad (7)$$

For $\alpha = 1$, Lévy distribution can be simulated using the following formula

$$\zeta = \frac{2\sigma}{\pi}\left[(\frac{\pi}{2} + \beta U)\tan(U) - \beta \ln(\frac{\frac{\pi}{2} W \cos(U)}{\frac{\pi}{2} + \beta U})\right] + \mu. \quad (8)$$

Here $U$ and $W$ are independent random variables, with $U$ uniformly distributed on $(-\pi/2, \pi/2)$ and $W$ a standard exponential distribution. In Fig.2, the probability density functions $L_{\alpha,\beta}(\zeta;\sigma,\mu)$ with various stability indexes and asymmetric parameters are plotted, and some parameters are chosen by fixing $\mu = 0, D = 1.0$. Note that $L_{\alpha,\beta}(\zeta;\sigma,\mu)$ is symmetric for $\beta = 0$, and when $\alpha = 2$, $L_{\alpha,\beta}(\zeta;\sigma,\mu)$ is the well known Gaussian distribution. The heavy-tails are visible when $\alpha \neq 2$. Fig.2(b) illustrates that the changes of stability index $\alpha$ not only make tails heavier but also change the shapes of distribution, and when we decrease the stabilty index, $L_{\alpha,\beta}(\zeta;\sigma,\mu)$ becomes a very skewed distribution.

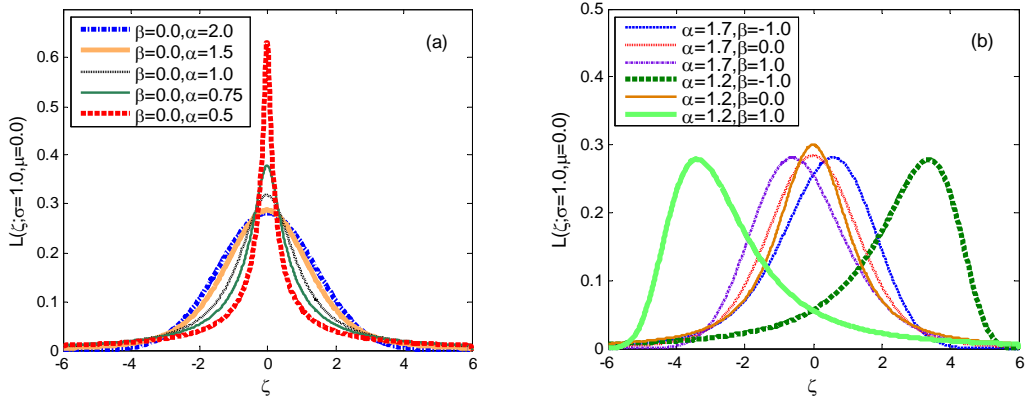

Fig.2. Probability density functions for Lévy distribution $L_{\alpha,\beta}(\zeta;\sigma,\mu)$ with different stability indexes (a) and asymmetry parameters (b).

## III. Stochastic Resonance and Simulation Results

This section is devoted to computing signal-to-noise ratio (SNR) which is used to measure SR phenomenon of the system (1). The significance of SR is the enhancement of output signal via tuning the noise intensity. This may be described as follows: At a lower noise level, the particle oscillates at the bottom of the potential wells for a long time and rarely switches between two potential wells. However, when we slowly increase the noise intensity, switching can be induced and the response of system becomes strongly nonlinear, this will cause the dropping of the output SNR. But after that when we modulate the coincidence of the noise and signal at a certain range, the output SNR will give rise to a local maximum (SR effect), and at bigger noise levels, the effect of the potential barrier gradually diminish, which will lead to less nonlinear behavior of response and the decrease of output SNR once again. Based on this observation, the SNR, as a function of noise intensity, is used to quantify the SR effect, and the maximum in the SNR would be detected by adjusting the noise intensity.

### A. Simulation Analysis

This part focuses on the numerical scheme to simulate the SNR function. As is known, the common definition of SNR is the ratio between the power spectral density of a signal and noise at the signal frequency. A few authors [1,2,36] have proposed several numerical methods of SNR. One method [1,11,19,21,25] is described as follows

$$SNR = \frac{2}{S_N(\Omega)} \lim_{\Delta\omega \to 0} \int_{\Omega-\Delta\omega}^{\Omega+\Delta\omega} S(\omega)d\omega, \tag{9}$$

where $\int_{\Omega-\Delta\omega}^{\Omega+\Delta\omega} S(\omega)d\omega$ represents the power carried by the signal, $S_N(\omega)$ estimates the background noise level, and $S(\omega) = \int_{-\infty}^{+\infty} e^{-i\omega\tau} \langle\langle x(t+\tau)x(t)\rangle\rangle d\tau$, with the inner brackets denoting the ensemble average over the realizations of the noise and outer brackets indicating the average over the input initial phase. Another method of computing SNR is in the following expression [36]

$$SNR = \frac{P_{signal}}{P_{noise}} = \frac{P - P_{noise}}{P_{noise}}, \tag{10}$$

where $P_{signal}$ and $P_{noise}$ represent the power spectrum of the signal and noise respectively, and $P$ is the total output power spectrum of system. In contrast to the above two methods, we calculate the SNR as the common logarithm of the power spectrum ratio between the signal $P_{signal}(\omega)$, and noise $P_{noise}(\omega)$, components [2]:

$$SNR = 10\log \frac{P_{signal}(\omega)}{P_{noise}(\omega)}. \tag{11}$$

By definition, SNR is measured in decibel (dB) units. In order to gain SNR of system, we let $\dot{x} = y$, and then the Eq.(1) is replaced by two first-order differential equations

$$\begin{cases} \dot{x} = y, \\ \dot{y} = -\gamma y - \frac{dV(x)}{dx} + F_1(t) + \eta(t), \end{cases} \tag{12}$$

and the discrete form has been taken as [22,26]:

$$\begin{cases} x_{n+1} = x_n + y_n \Delta t, \\ y_{n+1} = y_n + \left[-\gamma y_n + ax_n - bx_n^3 + F_1(n\Delta t)\right] \cdot \Delta t + \Delta t^{1/\alpha} \zeta, \end{cases} \tag{13}$$

where $\zeta$ is a Lévy distributed random number with stability index $\alpha$ and noise intensity $D$. Then the numerical solution of Eq.(1) is computed by using the fourth-order Runge-Kutta method. In practice, the noise intensity is very small, so $P_{signal}(\omega) = |Y(k_0)|^2$ can be used to estimate the signal's power spectrum, and $Y(k)$ is the fast Fourier transform (FFT) of the output signal sequence $\{x(n), n = 0, 1, \cdots, N-1\}$, i.e.,

$$Y(k) = \sum_{n=0}^{N-1} x(n) e^{-2\pi jkn/N}, \quad (14)$$

where $N$ is the data length of Lévy distribution sequence. Here we take $N = 2^{13}$. Due to the power spectrum of white noise changes gently and for a given signal, the signal frequency corresponding to the maximum of $|Y(k)|^2$ ($k = 0, \cdots, N-1$) doesn't change with the variation of the noise intensity, and owing to the corresponds of the signal frequency and integer $k_0$ in the discrete Fourier transform, so $k_0$ is defined as the $k$ where $|Y(k)|^2$ obtain the maximum value under small noise intensity, and with the increase of noise intensity, the value of $k_0$ will remain unchanged. The noise power spectrum $P_{noise}(\omega)$ is estimated in the signal bins by taking an average of the output power spectrum of system in neighboring bins (these bins contain noise only) $k_0 - M, \cdots, k_0 - 1, k_0 + 1, \cdots, k_0 + M$ for some integer $M$ [2],

$$P_{noise} = \frac{1}{2M} \sum_{i=1}^{M} (|Y(k_0 + i)|^2 + |Y(k_0 - i)|^2). \quad (15)$$

Here $M = 5$ defines a bandwidth to approximate the noise component at signal frequency. Especially, when we use the fourth-order Runge-Kutta method to get the numerical solution of Eq.(1), the heavy-tails, discontinuity and irregular jumps of Lévy noise may cause the numerical sample paths going to infinity rapidly with the decrease of the stability index $\alpha$. To solve this problem, some methods have been used, for example, Zhang and Song [11,25] imposed a constraint on the value of the solution $x(t)$. On the contrary, here we choose 15000 samples to get SNR, and fix the initial value of $x(0)$ and $\dot{x}(0)$ with zero. Through calculating the solutions of Eq.(1) for partial samples, it is clear that these solutions appear around $\pm 1$, and on these ground, we can set a critical value to $x(t)$. The critical value will be defined as $10^{13}$, and for one sample, if the solution of Eq.(1) exceeds the critical value, we will abandon it (i.e. deem it as an invalidate sample) and switch to next sample. The final result is the statistical average of all the effective samples.

To test the validity of our proposed numerical method, Fig.3 presents the results of SNR in the presence of Gaussian noise, which is a special case of Lévy noise with $\alpha = 2, \beta = 0$. Here three damping parameters $\gamma = (0.4, 0.5, 0.6)$ are chosen respectively, and the results are obtained through statistical average for 1000 samples. The results are roughly consistent with the literature [32], but due to the employment of different definition of SNR, the values of D-axis and SNR-axis induce a relative shift.

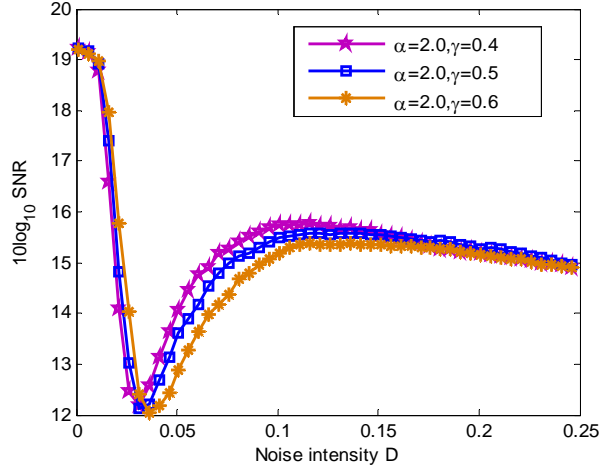

Fig.3. The SNR as the functions of the noise intensity for $a = 1$, $b = 1$, $w = \sqrt{2}/20$ and $A = 0.15$.

### B. Numerical Results

In the following, we use the above numerical method to simulate SNR function versus noise intensity $D$ for different parameters. The presence of the maximum in SNR will indicate the occurrence of SR in the system (1). The parameters are fixed as $a = 1, b = 1, w = \sqrt{2}/20, \beta = 0, \mu = 0$.

### 1. The variation of output signal and power spectrum with Lévy noise

First we investigate the effects of noise on the evolution of $x(t)$. Fig.4 shows the trajectories of input/output signal for various noise intensities with $A = 0.15$ ($A < \Delta V = 0.25$). In Fig.4 (a), we observe that the motion is confined in a single potential well with a small level noise, and some sharp spikes are clearly visible for the heavy tails of Lévy noise. From Fig.4 (b) we know that the decrease of noise intensity can cause the hopping of the particle between two potential wells, and this jump speed is found to accelerate with the increase of the noise intensity $D$. Fig.4 (a) and Fig.4 (b) illustrate that for a weak signal, noise intensity also may modulate the potential wells and proper noise intensity may lead to the SR effect.

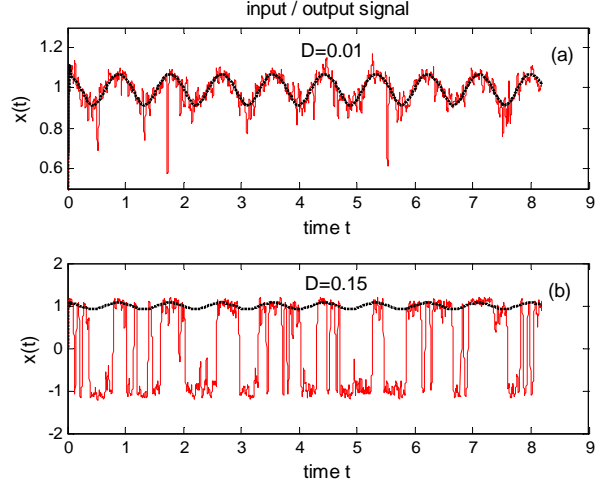

Fig.4. The trajectories of input (dashed black)/output (solid red) signal. The parameters are chose $A = 0.15$, $\alpha = 1.75$, $\gamma = 0.02$, $w = \sqrt{2}/20$ and the time step of the integration $\Delta t = 0.001$.

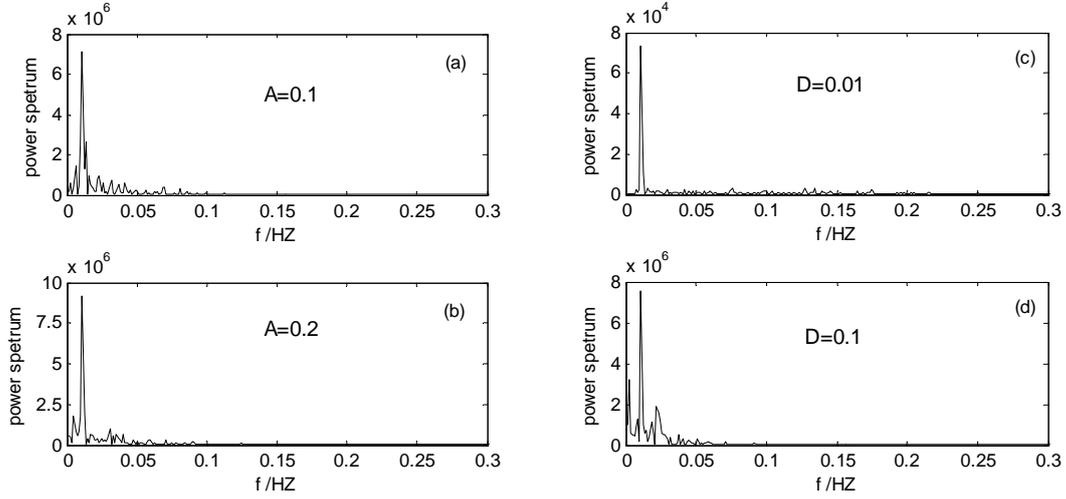

Fig.5. The power spectrum with different amplitudes (a) $A = 0.1$, (b) $A = 0.2$ for $D = 0.1$ and different noise intensities (c) $D = 0.01$, (d) $D = 0.1$ for $A = 0.15$. Other parameters of the system (1) are the same as in Fig. 4.

From section II, we know that the amplitude $A$ can modulate potential wells of the system (1), so suitable $A$ may also be beneficial to observe the SR. In Fig.5, we display the power spectrum for different values of amplitudes $A$ of periodic signal and noise intensities $D$. In all subfigures, the power spectrum of the system (1) on background of noise has a sharp peak at the driving frequency $f_0 = 0.01099$, which indicates that there is a stronger signal at frequency $f_0$. This peak value is a significant observation indicating the possibility of the occurrence of SR. Fig. 5 (a) and (b) show that the bigger amplitude $A(A < A_c = \sqrt{4/27})$ can increase the peak values of the power spectrum. Similarly, Fig. 5 (c) and (d) exhibit the increase of the noise intensity in a certain range also can enhance the peak values. However, when the noise intensity exceeds a critical value, this peak values will decrease as the noise intensity increases. These are the typical features of SR.

**2. The SNR functions with different parameters**

According to the above analysis, we try to find a value of noise intensity $D_{SR}$, at which the SR phenomenon occurs. In the following we simulate SNR functions with varying noise intensity $D$, stability index $\alpha$ and other parameters of the system (1), in order to quantify the SR phenomenon. The maximum of SNR in a particular regime of the noise intensity indicates the presence of the SR. For fixed amplitude $A$ ($A = 0.2 < \Delta V$), Fig. 6 shows the SNR versus the noise intensity $D$, with various damping parameters, and a non-monotonic behavior of SNR is observed. Fig. 6(a) exhibits the SNR function in the case of $\alpha = 1.75$, and Fig. 6(b) displays the situation of $\alpha = 1.9$. As can be seen in all subfigures, the values of SNR decrease with the increasing noise intensity $D$ at first, then begin to increase, and when the noise intensity reach to a critical value $D_{SR}$, the values of SNR achieve a maximum and after that decrease again. Under the different damping parameters, the non-monotonic behaviors of the SNR clearly indicate the occurrence of SR phenomenon. As increasing value of the damping parameter $\gamma$ from 0.02 to 0.7, the maximum of SNR drops and slightly shifts towards bigger values of noise intensity, and becomes more visible when $\gamma = 0.02$, thus $\gamma = 0.02$ is the optimal critical value to find SR. Comparing Fig. 6 (a) and (b), we observe that the decrease of stability index $\alpha$ weakens the maximum of SNR, and makes the SNR curves more rougher (i.e. the decreasing of stability index makes more and more samples' numerical integration paths escape to infinity and consequently reduces the numbers of valid samples), from this we conclude that a smaller stability index weakens the SR phenomenon, and more detailed discussions will be demonstrated in Fig.8.

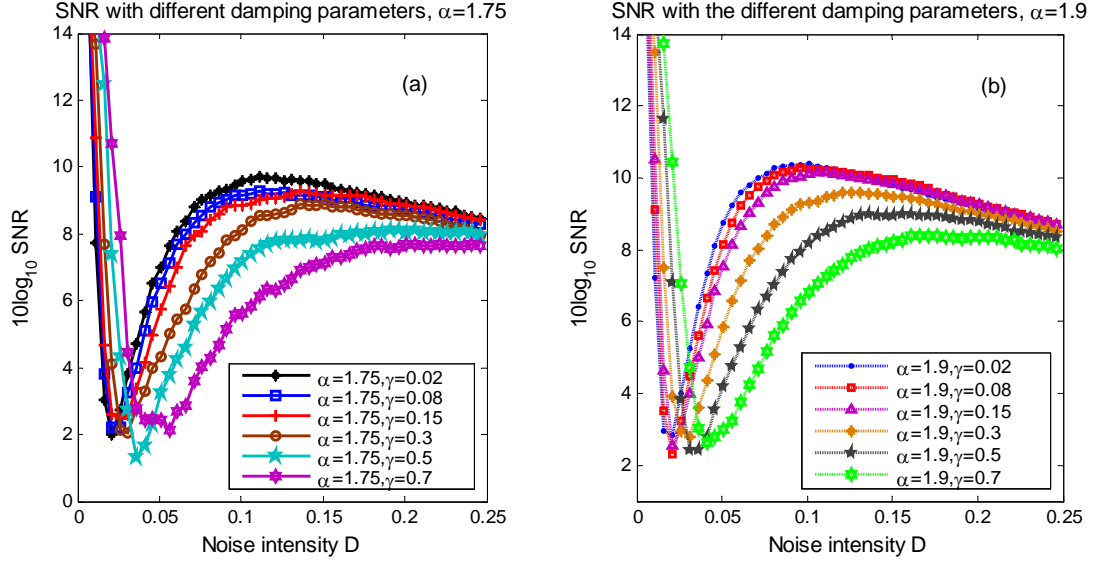

Fig.6. SNR as a function of the noise intensity $D$ with different stability index $\alpha$. (a) $\alpha = 1.75$, (b) $\alpha = 1.9$. The other system parameters are $a = 1$, $b = 1$, $A = 0.2$, $w = \sqrt{2}/20$ and $\Delta t = 0.001$.

The amplitude of external periodic signal also affect the power spectrum of output signal, so the optimal amplitude $A$ may lead to the optimal occurrence of the SR effect. Fig. 7 represents the SNR curves as a function of noise intensity $D$ for different modulation signals with amplitude $A = 0.15, 0.2, 0.25, 0.3, 0.35, 0.38$. All these amplitude values satisfy the condition $A < A_c = \sqrt{4/27}$. We note that the maximum value of the SNR decreases and shifts to larger noise level with the decreasing amplitude $A$. In fact, Fig. 7 also shows that a relatively smaller noise intensity has a negative influence in the system, but by tuning the noise intensity to $D_{SR}$, the system's output response will be strengthened, and then weakened while $D > D_{SR}$. This is a clear occurrence of SR.

For smaller amplitude $A$, the jumping of the particle between two potential wells demands bigger noise intensity, so the optimal noise intensity $D_{SR}$ becomes greater with the decreasing amplitude $A$. However, when we choose $A = 0.38$, which is close to the critical value $A_c$, the SNR could not decrease at first; on the contrary, it directly increases to maximum and afterwards decays with enhancing noise intensity. It indicates that for a stronger external signal, the particle can easily jump between two potential wells under lower noise, and larger amplitude $A$ is more beneficial to the SR effect.

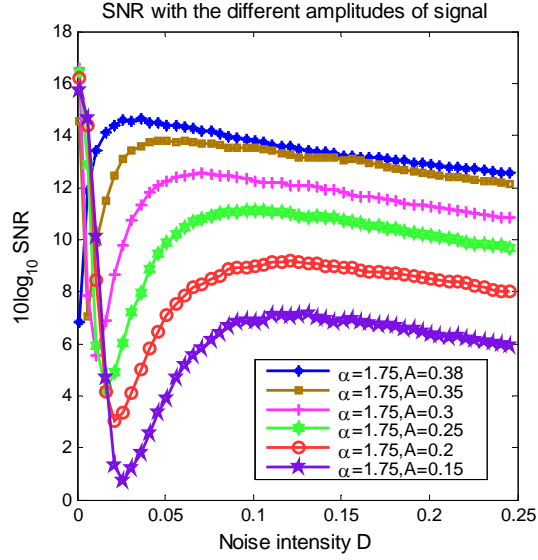

Fig.7. The SNR function as a function of the noise intensity $D$, for various amplitudes $A$ but for fixed $\gamma = 0.02$ and $\Delta t = 0.00095$. The other parameters are identical with Fig. 6.

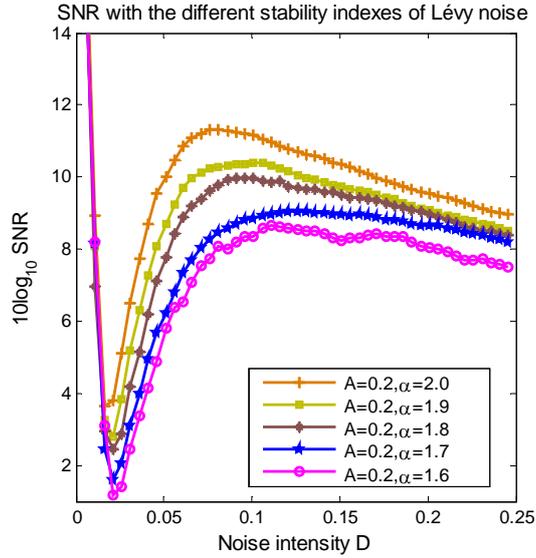

Fig.8. The SNR as a function of the noise intensity $D$, for different stability index $\alpha$ of Lévy noise, namely $\alpha = 1.9, 1.8, 1.7, 1.6$, but for fixed $\Delta t = 0.001$ and $\gamma = 0.02$. Simulation details are the same as in Fig.6.

Fig.6 demonstrates that the stability index $\alpha$ of Lévy noise can influence the SNR function, and further analysis of SNR function with different stability indexes $\alpha$ is presented in Fig.8. All SNR curves have the same trend with different $\alpha$, and the peak value of SNR drops and shifts to larger noise intensity with the decreasing of $\alpha$. This is because lower stability index deteriorates the output of the system (1). At the same time, for a fixed noise intensity $D$, the smaller index $\alpha$ results in heavier jumps of Lévy noise which makes the switching events between two potential wells more frequent, and the sharp spikes of trajectory $x(t)$ become more pronounced with smaller stability index. The figure implies the decrease of $\alpha$ weakens the occurrence of SR phenomenon.

## IV. Discussions and Conclusions

We have investigated the SR effect numerically for a second-order system driven by Lévy noise and used the signal-to-noise ratio (SNR) to quantify the stochastic resonance (SR) phenomenon. Our primary interest is to study the dependence of the SNR functions on different amplitudes of signal, stability indexes and other system parameters.

It is observed that the noise intensity and amplitude of external signal affect the system's output power spectrum that reaches a peak value at a constant frequency. The increase of the amplitude or noise intensity in a certain range further enhances this peak value, which illustrates that the appropriate noise intensity and amplitude lead to the optimal occurrence of the SR phenomenon. Additionally, for different system parameters, a non-monotonic behavior of SNR is shown and it is possible to find the optimal noise intensity value $D_{SR}$ at which the SR occurs. Meanwhile, the maximum of SNR drops and slightly shifts towards bigger noise intensity with the increase of the damping parameter. We also find that larger amplitude $A$ promotes the occurrence of SR. Finally, the SNR functions versus the noise intensity $D$ are exhibited for different values of stability index $\alpha$, and the results indicate that the decrease of stability index $\alpha$ leads to larger fluctuations and heavier tails of the noise, which causes the switching events between two potential wells more frequent and consequently weakens the SR phenomenon.

In conclusion, the SR effect induced by Lévy noise is observed in bistable second-order systems, and the influences of different parameters on the optimal SR phenomenon are carefully examined. We believe that the numerical method proposed in this paper provides the most effective way of understanding the mechanism of SR and offers new perspectives on what concerns the consideration of SR as a general phenomenon that might apply to diverse system.

## Acknowledgments

This work was supported by the NSF of China (Grant Nos. 10972181, 11102157), Program for New Century Excellent Talents in University, the Shaanxi Project for Young New Star in Science & Technology, NPU Foundation for Fundamental Research and New Faculty and New Area Project. We also thank Barbara Gentz and Ilya Pavlyukevich their valuable discussions.